\documentclass{desyproc}
\usepackage{braket}
\usepackage{color}

\begin{document}
%------------------------------------
\title{\vspace{-2.05cm}
\hfill{\small{DESY 14-166}}\\[1.27cm]
Hidden Photon Dark Matter Search with a Large Metallic Mirror}

%for single authors the superscripts are optional
\author{{\slshape Babette D\"obrich$^1$, Kai Daumiller$^2$, 
Ralph Engel$^2$, Marek Kowalski$^{1,3}$, 
Axel Lindner$^1$, Javier Redondo$^4$, Markus Roth$^2$}\\[1ex]
$^1$Deutsches Elektronen-Synchrotron (DESY), 22607 Hamburg and 15738 Zeuthen, Germany \\
$^2$Karlsruher Institut f\"ur Technologie (KIT), IKP, 76021 Karlsruhe, Germany \\
$^2$Humboldt Universit\"at, Institut f\"ur Physik, 12489 Berlin, Germany \\
$^4$Zaragoza University, Pedro Cerbuna 12,
E-50009 Zaragoza, Spain}

% if the proceedings are available online (e.g. at Indico)
% please enter the contribution ID or file_name below for the DOI
%\contribID{32}
\contribID{doebrich\_babette}

% TO THE CONFERENCE EDITORS: 
% please update the following information      
% before sending the template to the authors
% \confID{800}  % if the conference is on Indico uncomment this line
\desyproc{DESY-PROC-2014-XX}
\acronym{Patras 2014} % if you want the Acronym in the page footer uncomment this line
\doi  % if there is an online version we will register DOIs

\maketitle

\begin{abstract}
If Dark Matter is composed of hidden-sector photons that kinetically
mix with photons of the visible sector, then Dark Matter has a tiny 
oscillating electric 
field component. Its presence would lead to a small amount of visible radiation
being
emitted from a conducting surface,
with the photon frequency
given approximately by the mass of the hidden photon. Here, we report on
experimental efforts that have started recently to search
for such hidden photon Dark Matter in the (sub-)eV regime with a
prototype mirror for the Auger fluorescence detector
at the Karlsruhe Institute for Technology.
\end{abstract}

\section{Ultralight Dark Matter and the dish principle \label{sec:intro}}

In the literature there is no shortage of well-motivated candidates 
for cold Dark Matter (DM) particles. Without going into details of their respective theoretical
motivation, it is however clear that there is more experimental work needed in the search
for its ultra-light candidates below the eV regime: 
Although different detection schemes have been proposed,
only a few laboratory Dark Matter searches
are actively searching for low-mass particles such as QCD
Axions, see, e.g., recent progress of the Axion Dark 
Matter eXperiment (ADMX) \cite{Stern:2014wma}
and EDM-based techniques \cite{Budker:2013hfa}.

Considerations of general classes of ultra-light particles, dubbed `weakly
interacting slim particles' (WISPs) \cite{Baker:2013zta} have shown that such particles could make
up the Dark Matter in a rather large parameter space: 
particularly axion-like particles (ALPs) and massive hidden 
photons (HPs) \cite{Jaeckel:2013ija} can in principle
constitute all of the cold Dark Matter mainly through the misalignment mechanism which
is also invoked for Axions, see \cite{Arias:2012az}.
Whilst the viable parameter space for such ultra-light Dark Matter is
likely to be further constrained from cosmological observables,
ultimately laboratory experiments should be performed
to have certainty on its existence.

On the experimental side, set-ups like ADMX
are based on a {\it resonant} conversion of axions (and WISPs) and are thus ideal
to find extremely weakly coupled particles
in a rather narrow mass region. This is ideal for a QCD axion Dark Matter
search. For covering a wider mass-range, the search for ALP and HP Dark Matter with a 
spherical mirror has been recently proposed \cite{Horns:2012jf}:
Here the conversion is {\it not resonantly amplified} and thus the most immediate experimental
setups are
less sensitive with respect to the coupling (but have the advantage 
of broad-band frequency/mass coverage).

Let us recapitulate the idea of the `dish setup' for HP Dark Matter,
analogous considerations hold then for ALPs\footnote{From the 
experimental point of view, to look for ALP DM with this technique is rather
involved, since for a decent sensitivity the mirror has to be strongly magnetized 
with field strengths on the order of a few Tesla \cite{Horns:2012jf}. 
For the experimental setup at KIT described here, this
will likely not be possible.}: 
The relevant term for HP DM $\tilde{\gamma}$
is photon-to-hidden-photon coupling,
%$\chi F^{\mu \nu} X_{\mu \nu}$,
parameterized by the kinetic
mixing parameter $\chi$, see, e.g. \cite{Jaeckel:2013ija}. 
It eventually leads to electromagnetic power being emitted 
by a conducting surface (e.g. mirror)
at angular frequencies approximately corresponding to
the HP mass, $\omega \simeq m_{\tilde{\gamma}}$  \cite{Horns:2012jf}.
This is due to the presence of the HP DM together with the usual requirement that
for electric fields at the conducting surface $\vec{E}|_{\parallel}=0$.
To first order, photons are emitted perpendicular to 
the surface, with small corrections stemming from 
directionality of the DM inflow (which can be used to verify its DM origin).

To detect photons induced by this process, the advantage 
of using a spherical mirror is imminent:
photons from far away background sources impinging on the mirror will be focused in
the focal point $f=R/2$
whilst the Dark-Matter-induced photons 
will propagate to the center of the `mirror sphere'. 
There, a detector can be mounted.
A small off-set away from the center can be understood as follows:
Be 
$\vec{p}$ the momentum of the incoming DM, and $\vec{k}$ the
outgoing photon momentum, then 
$k_\parallel=p_\parallel$ 
along an infinitely extended surface because there is no boundary change (the approximation
is then valid as long as $\lambda$ is much smaller then the surface diameter).
With energy conservation $\vec{k}=\sqrt{m^2+|\vec{p}_\bot|^2}\vec{n} +\vec{p}_\parallel$,
with normal $\vec{n}$ to the surface.
As for the DM $|\vec{p}|\ll m$, the angular off-set 
of the signal away from the center of the `dish-sphere' is
$\psi\simeq |\vec{p}_\parallel|/m$ and the off-set on the detector is
$d_i \simeq \frac{p_i}{m} R$ when the
detector is at center $R$ and $i$ labels directions along the surface.

Nicely, thus, such a setup has a directional sensitivity \cite{Jaeckel:2013sqa}, which is easy to 
retrace within the common DM halo models.
E.g., assuming an isotropic velocity distribution
of the DM
with respect to the galactic frame,
a global off-set of the signal on the order of 
$\Delta d \sim \Delta v_{\rm detector} R$
%\sim 1 {\rm mm} \left(R/{\rm m} \right)$,
is expected due to the movement of the sun in the galactic rest frame
as well as a daily modulation on the same order of magnitude
(the yearly modulation is negligible due to the 
small velocity of the earth w.r.t. the sun).
Besides the signal-spot movement,
a likely velocity distribution $\Delta v_{\rm DM} \sim 10^{-3}$ 
of the DM leads to a broadening of the signal spot.
Ultimately, it is nice that this directional sensitivity 
can help to verify the Dark Matter nature of a signal.

\section{Prospective sensitivity with the KIT mirror}

\begin{figure*}[htbp]
\begin{tabular}{cc}
\begin{minipage}[t]{0.48\textwidth}
\begin{center}
\includegraphics[width=1\textwidth]{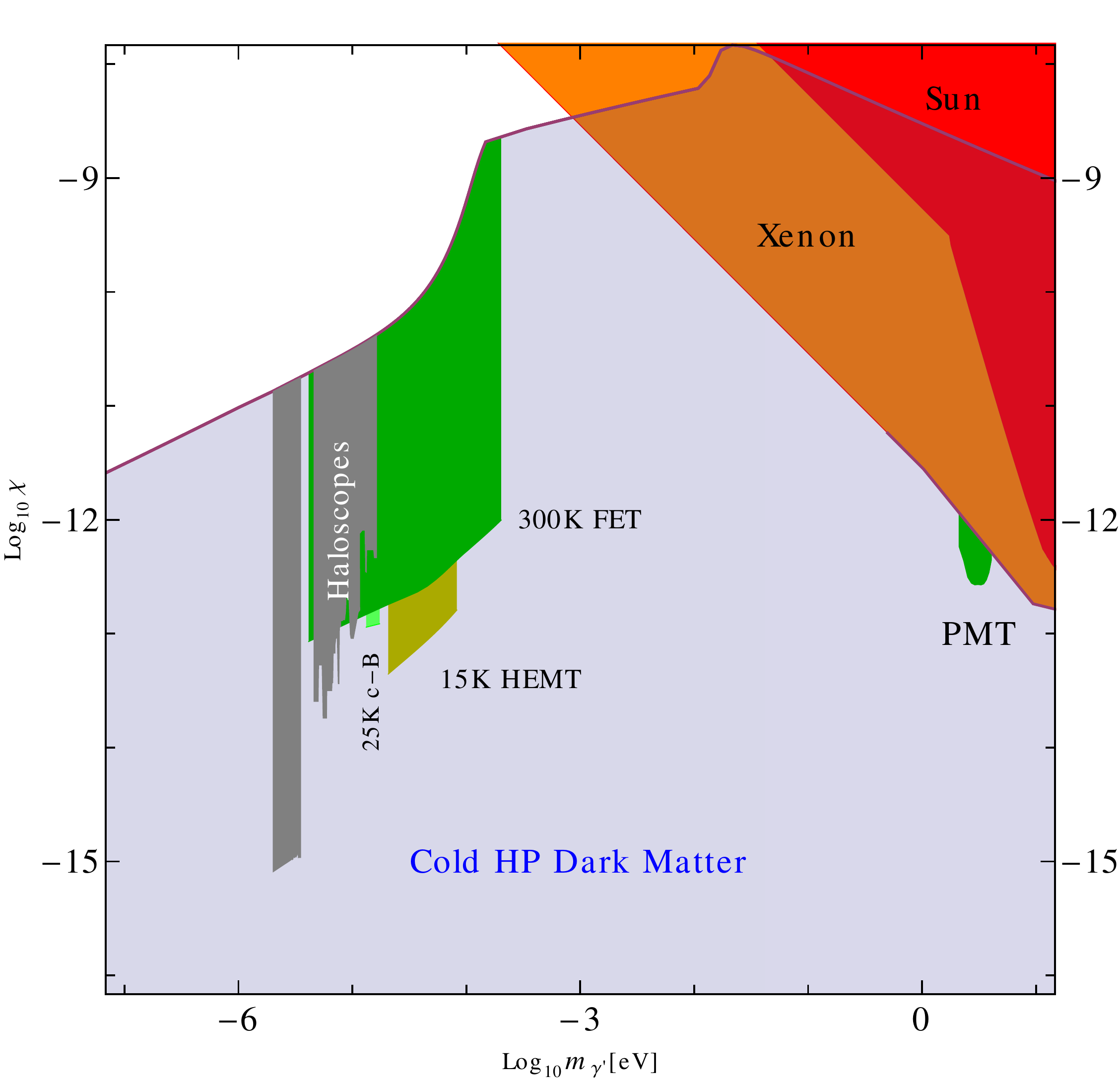}
\caption{Hidden photon DM parameter
space (blue) and exclusion regions (red/orange).
In green some parameter regions accessible with the metallic
mirror setup with different detector options. See text for details and \cite{Jaeckel:2013ija}
for a comprehensive review of the parameter space.
} 
\label{fig:excl}
\end{center}
\end{minipage}

%\hfill
&

\begin{minipage}[t]{0.5\textwidth}
\begin{center}
\includegraphics[width=1\textwidth]{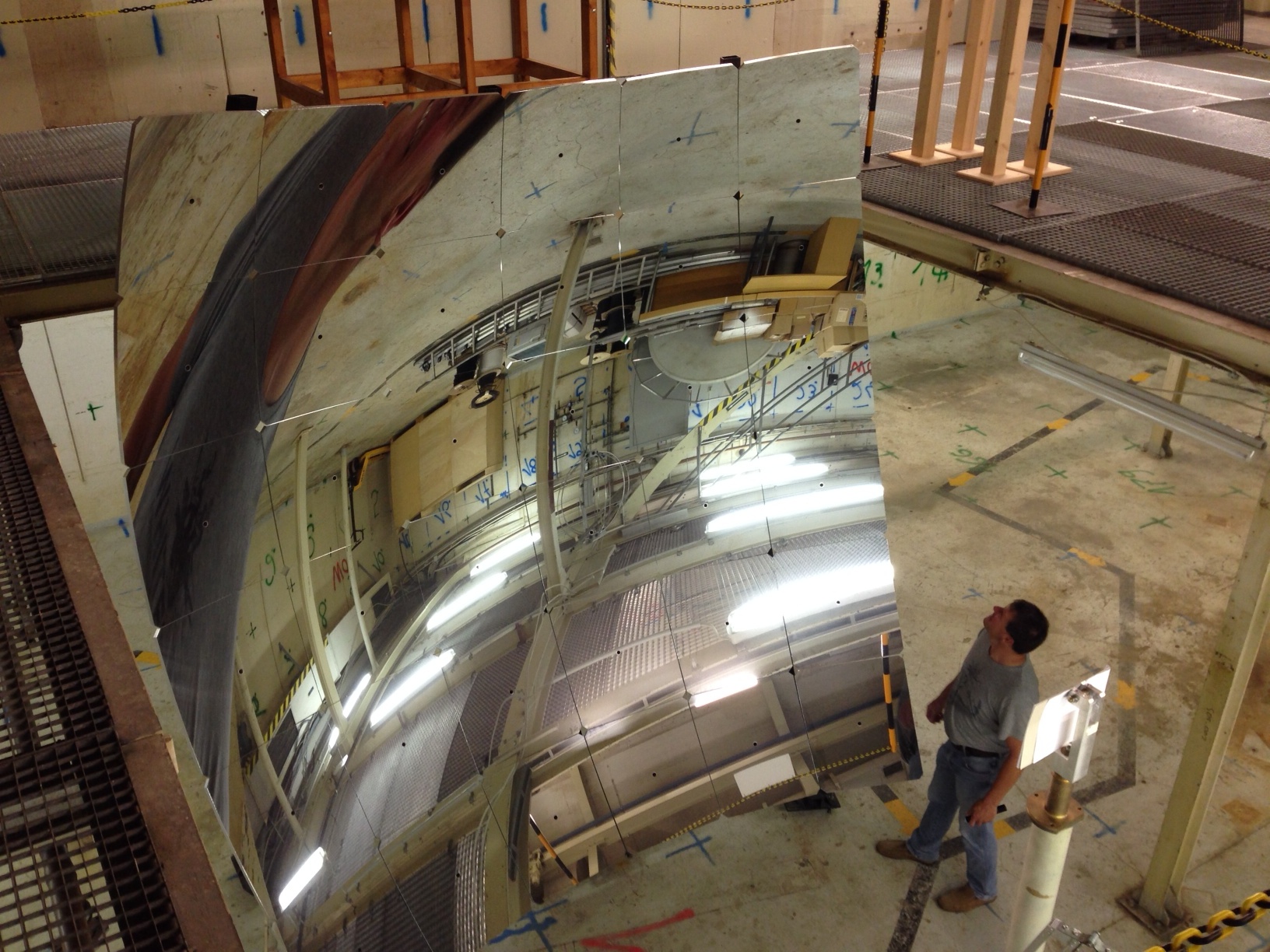} 
\caption{Spherical prototype mirror for AUGER housed at KIT (campus north). 
The grey post at the lower right hand side is the detector mount
located in the center of curvature.
}
\label{fig:dish}
\end{center}
\end{minipage}

\end{tabular}
\end{figure*}

The Pierre Auger Observatory uses two types of mirrors (coated glass
and coated aluminum)  \cite{Abraham:2009pm}. Both are are
segmented due to their rather large overall area
of $A\simeq 13 {\rm m}^2$, see Fig.~\ref{fig:dish}.
One prototype aluminum mirror for this experiment is kept at the Karlsruhe Institute for 
Technology (KIT).
As the mirrors are spherical with $R=3.4$m, the metallic mirror
is ideal for the Dark Matter search described above.
Assuming a Dark Matter density of  $\rho_{\rm CDM} \simeq 0.3 {\rm GeV}/{\rm cm^3}$
and assuming that HPs make up all of the Dark Matter 
the power emitted to the center is

\begin{equation}
 P= \braket{\alpha^2} \chi^2 \rho_{\rm CDM} A_{\rm dish}  \approx \chi^2 \ (1.87 \times 10^5 \ {\rm Watt}) \ ,
\end{equation}
where $\braket{\alpha^2}$ is a $\mathcal{O}(1)$-factor related to the polarization of the
HPs \cite{Horns:2012jf}, which we have taken to be one for simplicity.
As mentioned, the experimental advantage now is that the power is 
concentrated at the center $R$ of the `mirror sphere',
and not at the focal point $f=R/2$.

As benchmark number, one would like to probe the
parameter space below $\chi=3 \times 10^{-12} \ 1/m[{\rm eV}]$,
which is the limit inferred from the XENON10 experiment \cite{An:2013yua}, see the orange 
region labeled `Xenon' in Fig.~\ref{fig:excl}.

The setup described above is sensitive to all HP masses 
whose associated wavelength $\lambda=2\pi/m$ can be: 
1) detected by the sensor and 2) properly focused by the mirror (here we assumed $\lambda\ll R$ 
to use light-ray approximation and neglect diffraction, which would affect our estimates 
approximately below the mass range at which we cut  Fig.~\ref{fig:excl}).

For technological simplicity, measurements in the visible are a good starting point,
although their range is a limited. As an example, labeled `PMT' in Fig.~\ref{fig:excl}, 
we have plotted the sensitivity range
of a readily available\footnote{see, e.g.
\url{http://my.et-enterprises.com/pdf/9893_350B.pdf}.}, low-noise  ($\lesssim$1Hz) 
cooled PMT with $\sim25$\% quantum efficiency in the (300-500)nm regime
%in the visible 
at a SNR of 3 and 30h measurement time (assuming 
we are noise-limited by the detector, which is conceivable in the optical).

One can see that even this most simple and realistic setup is quickly sensitive to 
uncharted parameter space and with a set of PMTs, the near-infrared to UV range
can be explored down to $\chi\sim 10^{-13}$
(the overall coverage is a bit limited in the eV-range due to 
the strong bounds imposed by \cite{An:2013yua}).

Since the mirror is by default set up in a room with $\mathcal{O}$(m)-thick concrete walls
and further shielding can be constructed if required,
measurements down to the GHz-range can be envisioned and are sensitive to larger parameter
regions of HP DM. The gray regions in Fig.~\ref{fig:excl} 
indicate the exclusion set through a null-result of different
QCD axion haloscopes as described in Sect~\ref{sec:intro}. 

In green and yellow, again we plot parameter space accessible in principle to our set-up,
here within
a few minutes
in an idealized situation where we are limited by detector noise (if the
mirror has high reflectivity for the corresponding frequencies, its thermal emission should be low).
We sketch the accessible parameter region
using the Dicke radiometer equation
for a 25K c receiver at hand ($\sim 3.2-4.2$GHz) (lighter green).
For slightly higher frequencies we employ the noise figure
provided in \cite{Weinreb}. One sees that
even non-cryogenic options (300K FET, darker green) can cover a neat section of parameter space,
we also plot in lighter green the accessible region for a 15K HEMT (yellow).

Note the in the above considerations we have left out
implications of directionality discussed in Sect~\ref{sec:intro}. 
In the final analysis, the data sets
have to be evaluated in a particular DM model in which off-set and modulation
can be computed.
%(usually such mirrors are parabolic which...)

In summary, one sees that this rather simple setup offers many options
to look for HP Dark Matter. Fig.~\ref{fig:excl} just sketches the most immediate
options for this setting for good experimental conditions. If we are successful
in these first steps, measurements
in also in intermediate frequency ranges could be conceived.
In the following months, the results of ongoing background measurements
and budgetary considerations will determine our next steps.

\section{Summary}

Hidden Photons could constitute (part of) Dark Matter. To test this possibility,
cosmological guidance and laboratory
experiments are needed. A novel setup with a large metallic mirror
that can probe HP masses
in the $10^{-5}-10^{0}$ eV-regime down to kinetic mixing values of
$\chi \sim 10^{-13}$ is being set up at Karlsruhe.
This experiment can nicely complement other broadband efforts \cite{Graham:2014sha}
to probe even lower HP DM mass-scales with microwave cavities. 

\vspace{0.3cm}

\noindent
{\it The authors acknowledge the support
of the Helmholtz Alliance for Astroparticle Physics,
the lasting support of many `light-movers'
\cite{Redondo:2013hca}
and the kind
 collaboration
of 
I.~G. Irastorza,
J.~Jaeckel,
D.~Horns,
H.~Kr\"uger,
A.~Lobanov,
H.-J.~Mathes, 
A.~Ringwald,
J.-E.~v.~Seggern,
G.~Woerner
and D.~Veberic on different aspects of this experiment. 
BD would like to thank the PATRAS 2014 workshop organizers for a topical and
motivating conference.
}

% ****************************************************************************
% BIBLIOGRAPHY AREA
% ****************************************************************************

\begin{footnotesize}

\end{footnotesize}

% ****************************************************************************
% END OF BIBLIOGRAPHY AREA
% ****************************************************************************

\end{document}